\documentclass[a4paper]{jpconf}
\usepackage{graphicx,color}
\def\urlprefix{}
\def\url#1{}

\begin{document}
\title{Intersite electron correlations in a Hubbard model on inhomogeneous lattices}

\author{Nayuta Takemori$^{1}$, Akihisa Koga$^{1}$ and Hartmut Hafermann$^{2}$}

\address{${}^{1}$Department of Physics, Tokyo Institute of Technology, Meguro, Tokyo 152-8551, Japan 
${}^{2}$Mathematical and Algorithmic Sciences Lab, France Research Center, Huawei Technologies Co. Ltd., 92100 Boulogne-Billancourt, France}

\ead{takemori@stat.phys.titech.ac.jp}

\begin{abstract}
We study intersite electron correlations in the half-filled Hubbard model on square lattices with periodic and open boundary conditions by means of a real-space dual fermion approach. By calculating renormalization factors, we clarify that nearest-neighbor intersite correlations already significantly reduce the critical interaction. The Mott transition occurs at $U/t\sim 6.4$, where $U$ is the interaction strength and $t$ is the hopping integral. This value is consistent with the quantum Monte Carlo results. This shows the importance of short-range intersite correlations, which are taken into account in the framework of the real-space dual fermion approach.
\end{abstract}

\section{Introduction}
The successful development of dynamical mean field theory (DMFT)~\cite{Georges96} enables us to investigate a wide range of heavy fermion systems. In its real-space formulation (RDMFT), it has successfully been applied to inhomogeneous systems such as quasiperiodic systems~\cite{Takemori,Takemoriproc,Takemura}, fermionic cold atoms~\cite{cold1,cold2,cold3,cold4}, electron systems on surfaces~\cite{surface}, interfaces~\cite{interface1,interface2} and topological insulating systems~\cite{topological}.
However, this method does not take into account intersite correlation effects, which should be important at very low temperatures. 
Spatial correlations in homogeneous systems have therefore been treated by means of extensions of DMFT, e.g., cellular DMFT (CDMFT)~\cite{CDMFT}, dynamical cluster approximation (DCA)~\cite{DCA}, or diagrammatic extensions of DMFT~\cite{Kusunose,DGA,Rubtsov,Otsuki,Hartmut}.
Even though a generalization of CDMFT to inhomogeneous systems exists (the so-called I-CDMFT~\cite{I-CDMFT}), it is in general not clear how to treat inhomogeneous systems by means of CDMFT or DCA, due to the absence of translational symmetry. 
Diagrammatic extensions of DMFT, however, are naturally well suited to treat such systems~\cite{RDGA}.
Here we present a real-space formulation of the dual fermion approach (RDF), which includes spatial correlations beyond RDMFT. It allows us to study intersite electron correlations in inhomogeneous lattices.
To clarify how intersite electron correlations affect Mott physics, we study the Hubbard model on square lattices with periodic  and open boundary conditions.
By calculating renormalization factors, we discuss intersite correlations for the Mott transitions in these systems.

This paper is organized as follows. In Sec.~\ref{sec:model}, we introduce the model Hamiltonian. In Sec.~\ref{sec:method}, we explain our theoretical approach. We discuss how the intersite electron correlations affect the Mott transition point in Sec.~\ref{sec:results}.
A brief summary is given in Sec.~\ref{sec:sum}.

\section{Model}
\label{sec:model}
We consider the single-band Hubbard model on the square lattice~\cite{Imada}, which is given by the following Hamiltonian
\begin{equation}
H=-t\sum_{\langle i, j \rangle} (c^{\dagger}_{i \sigma} c_{j \sigma}+ h.c.)+U \sum_{i}n_{i \uparrow}n_{i \downarrow},
\end{equation}
where the summation is over nearest-neighbors $\langle i,j \rangle$, 
$c^{\dagger}_{i \sigma} (c_{i \sigma})$ is 
a creation (annihilation) operator of an electron 
at the $i$th site with spin $\sigma(= \uparrow , \downarrow)$ 
and $n_{i \sigma}= c^{\dagger}_{i \sigma}c_{i \sigma}$.
$t$ is the transfer integral between sites and 
$U$ is the Coulomb interaction. 
We discuss Mott transitions in the half-filled system at finite temperatures,
setting the chemical potential to $\mu=U/2$.

\section{Method}
\label{sec:method}
In this paper, we use 
the dual fermion approach, which has been developed recently~\cite{Kusunose,Rubtsov,Otsuki,Hartmut}. 
In the method, auxiliary fermions, so-called dual fermions ($f,f^{*}$) are introduced via a Hubbard-Stratonovich transformation,
and intersite correlations are treated by means of diagrammatic expansions.
The diagrammatic expansion is formalized by the dual propagator
$[\hat{G}_{\omega\sigma}^{\rm{d}}]_{ij}=  -\langle{f_{i\omega\sigma}f_{j\omega\sigma}^{*}}\rangle$ , where $\omega$ denotes fermionic Matsubara frequency $i\omega_n = (2n + 1)\pi/\beta$.

Since the dual fermion approach has originally been formulated in momentum space, inhomogeneous systems cannot be treated in this framework. Therefore, we introduce the real-space formulation of this method. Here we only give an outline, a detailed derivation will be presented elsewhere.
In this formulation, the concerned system quantities are represented as $N\times N$ matrices (denoted by a ``hat''), where $N$ is the total number of sites.
For example, the system bare dual Green's function is denoted $[\hat{G}^{\rm d,0}_{\rm syst}]_{\omega}=-\hat{g}_{\omega}[\hat{g}_{\omega}+(\hat{\Delta}_{\omega}-\hat{t})^{-1}]^{-1}\hat{g}_{\omega}$, 
where $\hat{\Delta}_{\omega}$ denotes the hybridization function, $\hat{g}_{\omega}$ is the impurity Green's function and $[\hat{t}]_{ij}=t\delta_{\langle i,j \rangle}$.
Correlations between any two sites are straightforwardly included through the corresponding diagram connecting these two sites.
By considering first-order and second-order diagrams, the dual self-energy ${\hat{\Sigma}^{\rm d}}$ can be written
\begin{eqnarray}
\hat{\Sigma}^{\rm {d}}_{ij}=\hat{\Sigma}^{\rm{d(1)} }_{i}\delta_{ij} +\hat{\Sigma}^{\rm{d}(2)}_{ij}.
\end{eqnarray}
Here, we omit frequency labels for clarity.
We note that in the paper, we only consider intersite correlations between nearest neighbor sites $(\hat{\Sigma}^{\rm{d}(2)}_{ij} \sim \hat{\Sigma}^{\rm {d}(2)}_{ i,j }\delta_{\langle i,j \rangle})$, which we expect to be dominant (correlations of larger distances are straightforwardly included). Therefore, our method does not reduce to the conventional dual fermion approach even in the homogeneous system. Nevertheless, we can discuss the effect of intersite correlations beyond DMFT.
Then the system self-energy is given as
\begin{eqnarray}
\label{eq:sigma}
\hat{\Sigma}_{\rm{syst}} = \hat{\Sigma}_{\rm{imp}}^{i} + [(\hat{1} + \hat{\Sigma}^{\rm{d}}{\hat{g}_{\rm{imp}}} )^{-1} \hat{\Sigma}^{\rm{d}}].
\end{eqnarray}
The self-consistency condition for the real-space dual fermion approach is the same as that in conventional dual fermion approach,
\begin{eqnarray}
\hat{G^{\rm{d}}}_{ii} = 0.
\end{eqnarray}
Since the off-diagonal component of the dual self-energy $\hat{\Sigma}^{\rm d}$ originate from the vertex function, the real-space dual fermion approach reduces to RDMFT in the weak coupling limit, or when no diagrammatic corrections are taken into account. In these cases, the system self-energy is diagonal and the dual self-energy is equal to zero. 

In order to discuss Mott transitions at finite temperatures,
we calculate the following quantities,
\begin{equation}
\label{eq:z}
z_i=\left[ 1 - {\rm Im} \hat{\Sigma}_{\rm syst}^{ii} (i\omega_0)/\omega_0  \right]^{-1}.
\end{equation} 
They can be regarded as renormalization factors at finite temperatures.
In the following, we use the transfer integral $t$ as the unit of energy.
We treat the two dimensional square lattice with $4 \times 4$ sites with periodic  and open boundary conditions to discuss the effect of intersite correlations.

\section{Results}
\label{sec:results}

%%%%%%%%%%%%%%%%%%%%%%%%%%%%%%%%%%%%%%%%%%%%%%%%%%%%%%%%%%%
\begin{figure}[htb]
\includegraphics[width=38pc]{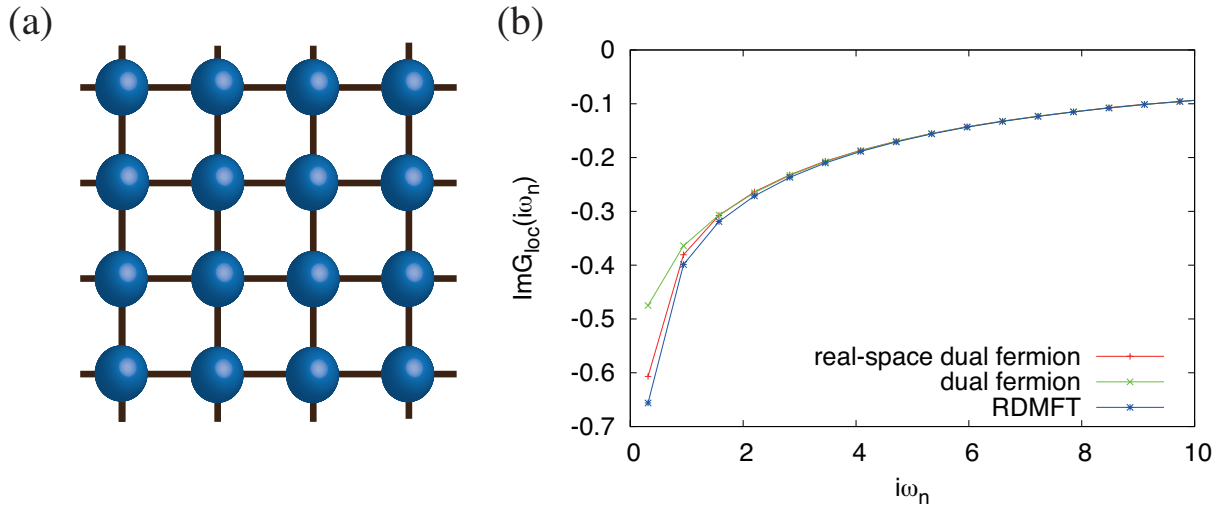}
\caption{\label{fig:1}
(a) Square lattice with periodic boundary condition.\\
(b)The imaginary part of the local Green's function $G_{\rm loc}(i \omega_n)$ for $U=4.0$ at $T=0.1$ obtained by RDMFT, dual fermion approach and real-space dual fermion approach. 
The unit of energy is set to be $t$.
}
\end{figure}
%%%%%%%%%%%%%%%%%%%%%%%%%%%%%%%%%%%%%%%%%%%%%%%%%%%%%%%%%%

We consider electron correlations in the system with periodic boundary condition (see Fig. \ref{fig:1}(a)).
To clarify how intersite correlations are taken into account in the real-space dual fermion approach, we calculate the local Green's function at $U=4.0$.
For comparison, we also show in Fig. \ref{fig:1}(b) results obtained by the RDMFT and conventional dual fermion approach.
These three results are in good agreement with each other in the intermediate energy region and show the same high-energy asymptotic behavior.
On the other hand, they deviate in the low energy region. We find that our results are located between those obtained by RDMFT and dual fermion methods. This is because our real-space dual fermion method takes into account some (i.e., those between nearest neighbor sites) but not all intersite correlations beyond the simple RDMFT. These correlations lower the spectral weight at the Fermi level.
The real-space dual fermion result must approach the conventional dual fermion result when considering longer-range correlations. This is currently under consideration.
We note that even though the effect of the intersite correlations on the Green's function is small at these parameters, we will see below that their effect can be significant.

\begin{figure}[h]
\includegraphics[width=25pc]{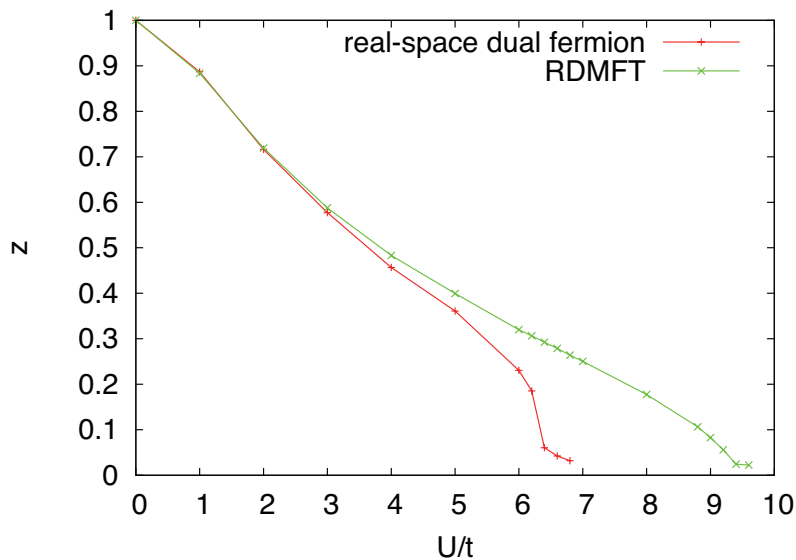}
\caption{\label{periodic}Renormalization factors in the system with periodic boundary condition obtained by means of the real-space dual fermion and RDMFT when $T=0.1$. The unit of energy is set to be $t$.}
\end{figure}

Figure \ref{periodic} shows the interaction dependence of renormalization factors in the system with periodic boundary conditions.
In the noninteracting case $U=0$, the normal metallic state is realized
with $z_i=1$.
By introducing the interaction, the renormalization factor at each site 
monotonically decreases, similarly as in the conventional Hubbard model~\cite{single}.
The small difference in the low-energy region in local Green's function obtained by RDMFT and real-space dual fermion approach, which is discussed already,  leads to the small supression in renormalization factors for  real-space dual fermion approach compared to RDMFT at $U=4.0$.
At last, the first-order phase transition occurs 
to the Mott insulating state at $U_c^{\rm periodic}$.
The critical interaction $U_c$ is deduced as $U_c^{\rm periodic}=9.4$ (RDMFT) and $6.4$ (real-space dual fermion approach).
In the homogeneous system, the transition point $U_c^{\rm periodic}\sim 6.5$ has been obtained precisely by the dual fermion approach~\cite{Hartmut}, $U_c^{\rm periodic}=6.5$ by DCA~\cite{Werner_private} and $U_c^{\rm periodic}\sim 6$ by CDMFT~\cite{square_CDMFT} and quantum Monte Carlo simulation~\cite{squareQMC}. Since the dominant correlations are between nearest-neighbors, finite size effects are expected to be small.
We conclude that the real-space dual fermion approach properly takes intersite correlations into account and gives us an improved picture of the Mott transition.

\begin{figure}[h]
\includegraphics[width=38pc]{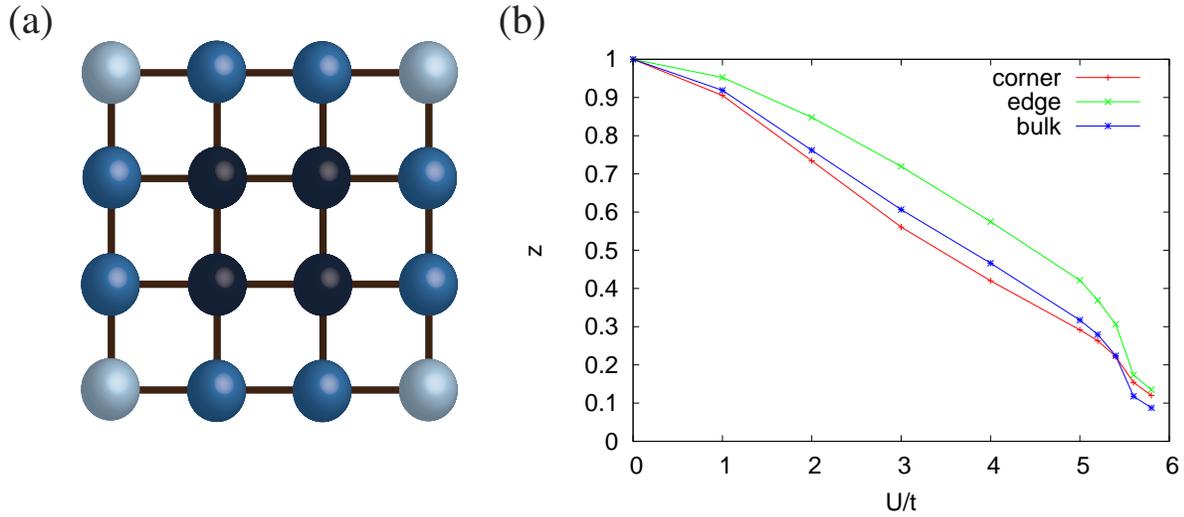}
\caption{\label{fig:3}
(a)Square lattice with open boundary condition.
Three independent sites are depicted in different gray scales.\\
(b)Renormalization factors in the system with open boundary condition obtained by means of the real-space dual fermion  when $T=0.1$. The unit of energy is set to be $t$.\\
}
\end{figure}

Next, we apply the approach to the Hubbard model with open boundary conditions (see Fig. \ref{fig:3}(a)) as an example of inhomogeneous lattices.
In this system with $4 \times 4$ sites, there exist three independent sites: the corner, edge, and bulk, as shown in Fig. \ref{fig:3}(a).
Figure \ref{fig:3}(b) shows the interaction dependence of renormalization factors for these sites.
It is found that these quantities are split into three lines in an interacting case.
This fact indicates that a site-dependent renormalization appears.
On the other hand, the Mott ``transition'' occurs simultaneously for all sites. Due to the open boundary conditions the critical value is lower and the transition point is deduced as $U_c^{\rm open}=8.6$ (RDMFT; not shown) and $5.6$ (real-space dual fermion approach), where the renormalization factor have jump singularities.
By comparing with the RDMFT results, we find a suppression of renormalization factors, which is similar to the homogeneous system.
Therefore, we conclude that real-space dual fermion approach can take intersite correlations into account correctly even in inhomogeneous system.

\section{Summary}
\label{sec:sum}
In summary, we have investigated the half-filled Hubbard model on square lattices with periodic  and open boundary conditions by means of real-space dual fermion approach.
By calculating local Green's function, we have studied how nonlocal correlations affect the local quantities. 
Moreover, we have discussed the Mott transition in this system to clarify that real-space dual fermion approach successfully captures 
the suppression of the renormalization factor. This leads to a Mott transition point with a relatively low critical Coulomb interaction compared to RDMFT.
The Mott transition point is close to that obtained by means of the quantum Monte Carlo method in the homogeneous system. Therefore, we can say that our real-space dual fermion method therefore takes intersite correlations properly into account, and the nearest-neighbor correlations are dominant in the system. 
This new method allows to study intersite electron correlation effects in various other inhomogeneous systems, such as cold atoms in a trapping potential, nanosystems, impurities, interfaces and surfaces, topological insulators and quasiperiodic lattices, which is now under consideration.

\ack
The authors would like to thank J. Otsuki and P. Werner for valuable discussions. 
This work was partly supported by JSPS KAKENHI Grant No. 15J12110 (N. T.) and  25800193 (A. K.).
Part of the computations was carried out on TSUBAME2.0
at Global Scientific Information and the Computing Center
of Tokyo Institute of Technology and at the Supercomputer
Center at the Institute for Solid State Physics, the University of
Tokyo.
The simulations have been performed using some ALPS libraries~\cite{alps2}.


\section*{References}
\begin{thebibliography}{99}
\bibitem{Georges96}
Georges A, Kotliar G, Krauth W and Rozenberg M~J 1996 {\em Rev. Mod. Phys.\/}
  {\bf 68}(1) 13--125
  \urlprefix\url{http://link.aps.org/doi/10.1103/RevModPhys.68.13}

\bibitem{Takemori}
Takemori N and Koga A 2015 {\em Journal of the Physical Society of Japan\/}
  {\bf 84} 023701 \urlprefix\url{http://dx.doi.org/10.7566/JPSJ.84.023701}

\bibitem{Takemoriproc}
Takemori N and Koga A 2015 {\em Journal of Physics: Conference Series\/} {\bf
  592} 012038 \urlprefix\url{http://stacks.iop.org/1742-6596/592/i=1/a=012038}

\bibitem{Takemura}
Takemura S, Takemori N and Koga A 2015 {\em Phys. Rev. B\/} {\bf 91}(16) 165114
  \urlprefix\url{http://link.aps.org/doi/10.1103/PhysRevB.91.165114}

\bibitem{cold1}
Helmes R~W, Costi T~A and Rosch A 2008 {\em Phys. Rev. Lett.\/} {\bf 100}(5)
  056403 \urlprefix\url{http://link.aps.org/doi/10.1103/PhysRevLett.100.056403}

\bibitem{cold2}
Snoek M, Titvinidze I, T\H{o}ke C, Byczuk K and Hofstetter W 2008 {\em New
  Journal of Physics\/} {\bf 10} 093008
  \urlprefix\url{http://stacks.iop.org/1367-2630/10/i=9/a=093008}

\bibitem{cold3}
Koga A, Higashiyama T, Inaba K, Suga S and Kawakami N 2008 {\em Journal of the
  Physical Society of Japan\/} {\bf 77} 073602 
  \urlprefix\url{http://dx.doi.org/10.1143/JPSJ.77.073602}

\bibitem{cold4}
Koga A, Higashiyama T, Inaba K, Suga S and Kawakami N 2009 {\em Phys. Rev. A\/}
  {\bf 79}(1) 013607
  \urlprefix\url{http://link.aps.org/doi/10.1103/PhysRevA.79.013607}

\bibitem{surface}
Potthoff M and Nolting W 1999 {\em Phys. Rev. B\/} {\bf 59}(4) 2549--2555
  \urlprefix\url{http://link.aps.org/doi/10.1103/PhysRevB.59.2549}

\bibitem{interface1}
Okamoto S and Millis A~J 2004 {\em Phys. Rev. B\/} {\bf 70}(24) 241104
  \urlprefix\url{http://link.aps.org/doi/10.1103/PhysRevB.70.241104}

\bibitem{interface2}
Okamoto S and Millis A 2004 {\em Nature\/} {\bf 428} 630--633
  \urlprefix\url{http://www.scopus.com/inward/record.url?eid=2-s2.0-1942421142&partnerID=40&md5=25c7747debc895d6cc3155cc79d50b53}

\bibitem{topological}
Tada Y, Peters R, Oshikawa M, Koga A, Kawakami N and Fujimoto S 2012 {\em Phys.
  Rev. B\/} {\bf 85}(16) 165138
  \urlprefix\url{http://link.aps.org/doi/10.1103/PhysRevB.85.165138}

\bibitem{CDMFT}
Kotliar G, Savrasov S~Y, P\'alsson G and Biroli G 2001 {\em Phys. Rev. Lett.\/}
  {\bf 87}(18) 186401
  \urlprefix\url{http://link.aps.org/doi/10.1103/PhysRevLett.87.186401}

\bibitem{DCA}
Hettler M~H, Tahvildar-Zadeh A~N, Jarrell M, Pruschke T and Krishnamurthy H~R
  1998 {\em Phys. Rev. B\/} {\bf 58}(12) R7475--R7479
  \urlprefix\url{http://link.aps.org/doi/10.1103/PhysRevB.58.R7475}

\bibitem{Kusunose}
Kusunose H 2006 {\em Journal of the Physical Society of Japan\/} {\bf 75}
  054713 \urlprefix\url{http://dx.doi.org/10.1143/JPSJ.75.054713}

\bibitem{DGA}
Toschi A, Katanin A~A and Held K 2007 {\em Phys. Rev. B\/} {\bf 75}(4) 045118
  \urlprefix\url{http://link.aps.org/doi/10.1103/PhysRevB.75.045118}

\bibitem{Rubtsov}
Rubtsov A~N, Katsnelson M~I and Lichtenstein A~I 2008 {\em Phys. Rev. B\/} {\bf
  77}(3) 033101
  \urlprefix\url{http://link.aps.org/doi/10.1103/PhysRevB.77.033101}

\bibitem{Otsuki}
Otsuki J, Hafermann H and Lichtenstein A~I 2014 {\em Phys. Rev. B\/} {\bf
  90}(23) 235132
  \urlprefix\url{http://link.aps.org/doi/10.1103/PhysRevB.90.235132}

\bibitem{Hartmut}
Hafermann H 2010 {\em Numerical Approaches to Spatial Correlations in Strongly
  Interacting Fermion Systems\/} (Cuvillier Verlag, G\"ottingen)

\bibitem{I-CDMFT}
Charlebois M, S\'en\'echal D, Gagnon A~M and Tremblay A~M~S 2015 {\em Phys.
  Rev. B\/} {\bf 91}(3) 035132
  \urlprefix\url{http://link.aps.org/doi/10.1103/PhysRevB.91.035132}

\bibitem{RDGA}
Valli A, Sangiovanni G, Gunnarsson O, Toschi A and Held K 2010 {\em Phys. Rev.
  Lett.\/} {\bf 104}(24) 246402
  \urlprefix\url{http://link.aps.org/doi/10.1103/PhysRevLett.104.246402}

\bibitem{Imada}
Imada M, Fujimori A and Tokura Y 1998 {\em Rev. Mod. Phys.\/} {\bf 70}(4)
  1039--1263 \urlprefix\url{http://link.aps.org/doi/10.1103/RevModPhys.70.1039}

\bibitem{single}
Rozenberg M~J, Chitra R and Kotliar G 1999 {\em Phys. Rev. Lett.\/} {\bf
  83}(17) 3498--3501
  \urlprefix\url{http://link.aps.org/doi/10.1103/PhysRevLett.83.3498}

\bibitem{Werner_private}
Werner P private communication

\bibitem{square_CDMFT}
Park H, Haule K and Kotliar G 2008 {\em Phys. Rev. Lett.\/} {\bf 101}(18)
  186403 \urlprefix\url{http://link.aps.org/doi/10.1103/PhysRevLett.101.186403}

\bibitem{squareQMC}
Veki\ifmmode~\acute{c}\else \'{c}\fi{} M and White S~R 1993 {\em Phys. Rev.
  B\/} {\bf 47}(2) 1160--1163
  \urlprefix\url{http://link.aps.org/doi/10.1103/PhysRevB.47.1160}

\bibitem{alps2}
Bauer B, Carr L~D, Evertz H~G, Feiguin A, Freire J, Fuchs S, Gamper L,
  Gukelberger J, Gull E, Guertler S, Hehn A, Igarashi R, Isakov S~V, Koop D, Ma
  P~N, Mates P, Matsuo H, Parcollet O, Paw{\l}owski G, Picon J~D, Pollet L,
  Santos E, Scarola V~W, Schollw{\"o}ck U, Silva C, Surer B, Todo S, Trebst S,
  Troyer M, Wall M~L, Werner P and Wessel S 2011 {\em Journal of Statistical
  Mechanics: Theory and Experiment\/} {\bf 2011} P05001
  \urlprefix\url{http://stacks.iop.org/1742-5468/2011/i=05/a=P05001}
\end{thebibliography}
\end{document}